\documentclass[12pt]{iopart}

\usepackage{iopams}  
\usepackage{graphicx}
\usepackage{subfig}
\usepackage{cite}
\begin{document}

\title{One dimensional magneto-optical compression of a cold CaF molecular beam}

\author{Eunmi Chae $^{1,2}$ \footnote{Present address: Photon Science Center, School of Engineering, the University of Tokyo, Japan 113-8656}, Loic Anderegg $^{1,2}$, Benjamin L Augenbraun $^{1,2}$, Aakash Ravi $^{1,2}$, Boerge Hemmerling $^{1,2}$ \footnote{Present address: Department of Physics, University of California, Berkeley, California 94720, USA}, Nicholas R Hutzler $^{1,2}$, Alejandra L Collopy $^{3}$, Jun Ye $^{3}$, Wolfgang Ketterle $^{2,4}$, John M Doyle $^{1,2}$} 

\address{$^{1}$Department of Physics, Harvard University, Cambridge, MA 02138, USA}
\address{$^{2}$Harvard-MIT Center for Ultracold Atoms, Cambridge, MA 02138, USA}
\address{$^{3}$JILA, National Institute of Standards and Technology and University of Colorado, Boulder, CO 80309, USA }
\address{$^{4}$Department of Physics, Massachusetts Institute of Technology, Cambridge, MA 02139, USA }

\ead{eunmi@cua.harvard.edu}

\begin{abstract}

We demonstrate with a RF-MOT the one dimensional, transverse magneto-optical compression of a cold beam of calcium monofluoride (CaF).
By continually alternating the magnetic field direction and laser polarizations of the magneto-optical trap, a photon scattering rate of $2\pi \times$0.4 MHz is achieved.
A 3D model for this RF-MOT, validated by agreement with data, predicts a 3D RF-MOT capture velocity for CaF of 5 m/s.

\end{abstract}

\maketitle

\section{Introduction}
Molecules are intriguing candidates for the study of fundamental symmetry violation, quantum simulation of strongly-correlated Hamiltonians, and the creation of new quantum information systems that take advantage of internal molecular degrees of freedom \cite{Carr2009, Baron2014, DeMille2008, Baranov2012, Buchler2007, Micheli2006, Yan2013b, Andre2006, DeMille2002, Rabl2006}. 
In addition, active research is on-going to create cold molecules for new studies in order to understand the full role of quantum mechanics in chemical reactions \cite{Ni2010h, Ospelkaus2010}.
The main difficulty in pursuing these goals is to controllably prepare molecules in single quantum states, and to achieve long interaction times (typically achieved via trapping).
Even the simplest molecules, diatomic molecules, have complex enough internal structure that, at room temperature, typically tens of thousands of states are thermally populated. 
Only very recently have molecular quantum states been controlled in a way to allow for quantum state-dependent chemical reactions \cite{Ni2010h, Ospelkaus2010}.
In that work, ultracold bi-alkali molecules were produced in a single quantum state by combining two ultracold alkali atoms, taking advantage of the mature, powerful tools for atom cooling \cite{Ni2008, Moses2015, Park2015a, Takekoshi2014, Molony2014, Guo2016}.
However, this technique is so far limited to a specific subset of molecules, typically bialkalis in the $^1\Sigma$ state.
Extension of these methods to more chemically diverse species is a formidable challenge.

To fully utilize the powerful diversity of molecules, new methods to produce cold and ultracold molecules (e.g.~those with electron spin degree of freedom) are desirable. 
A crucial step towards achieving lower temperatures is trapping, which in itself allows for further cooling methods to be applied.
Major efforts are currently ongoing to trap molecules using electric, magnetic, and optical forces \cite{Hummon2011c, Bethlem2000, VandeMeerakker2005h, Hoekstra2007e, Zeppenfeld2012a, Prehn2016a, Weinstein1998, Ye:07, Campbell2007, Lu2014, Stuhl2012e, Stoll2008, Riedel2011}. 
One promising option is a magneto-optical trap (MOT), the workhorse tool of optical cooling and confinement for atoms.
Despite thorough study and understanding of atomic MOTs, the additional internal structure present in molecules has made it difficult to produce a molecular MOT, mostly due to the lack of closed cycling transitions.
It is possible to overcome this difficulty by using molecules with sufficiently diagonal Franck-Condon factors (FCFs), as was proposed in references \cite{Rosa2004, Stuhl2008}.
Recently, 2D and 3D MOTs for molecules have been realized with YO and SrF \cite{Hummon2013, Barry2014, McCarron2015, Norrgard2016a}.
This scheme, although limited to molecules with strong optical transitions and highly diagonal FCFs, represents a powerful step toward broadening the scope of molecules that can be brought to the ultracold regime, including those of interest for new schemes in quantum simulation \cite{Micheli2006}.

CaF is a prototypical molecule for cold and ultracold applications and has been studied extensively, including its collisional properties \cite{Wall2008, Devlin2015, Maussang2005, Lim2015, Quemener2015}. 
It has an unpaired outer shell electron, strong laser cooling transitions, reasonably diagonal FCFs, and a relatively low mass (important for efficient beam slowing and MOT capture).
It is a molecule that is well suited for laser cooling.
Using the $X(v=0) - A(v'=0)$ transition at 606 nm with a linewidth of $2\pi \times 8.29$ MHz \cite{Wall2008}, about 10$^5$ photons can be scattered with 2 vibrational repump lasers, before falling into higher vibrational states \cite{Pelegrini2005}.
Recent theoretical work indicates that CaF is a good candidate for sympathetic/evaporative cooling to reach temperatures in the microkelvin regime, a necessary step toward quantum degeneracy \cite{Lim2015, Quemener2015}. 
Once an ensemble of ultracold CaF molecules is prepared, its large electric dipole moment and spin degree of freedom will expand the Hamiltonians that can be simulated, including spin-lattice models \cite{Micheli2006}.

Here we report a demonstration of magneto-optical compression of a CaF molecular beam.
This is a key milestone towards loading CaF into a 3D MOT.
This work verifies all of the processes and technology necessary for magneto-optical trapping of CaF, identifies an important MOT-limiting feature of CaF due to level crossings, and provides crucial data for validating RF-MOT loading models.
Our results also point towards the optimal parameters for achieving a RF-MOT with a maximum number of trapped CaF molecules.

\section{Methods}

\begin{figure}
\begin{center}
\includegraphics[width=\textwidth]{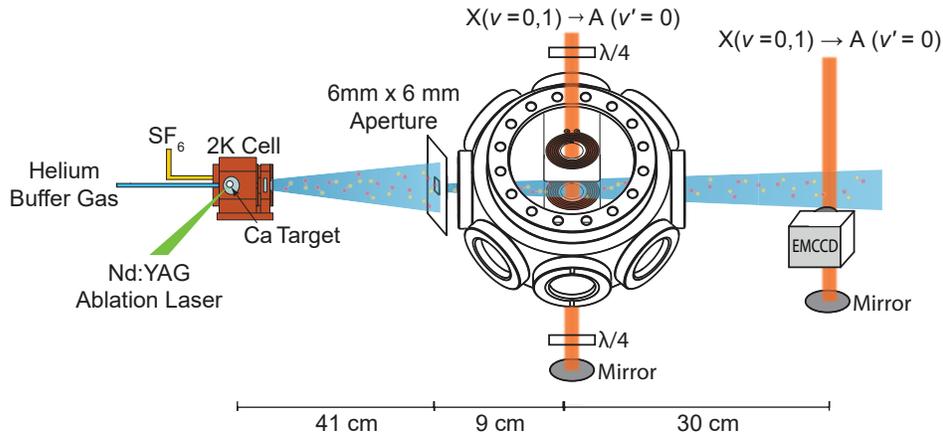}
\caption{Experimental setup (not to scale). The cell where molecules are generated is cooled by a pumped liquid $^4$He bath. An aperture of 6 mm $\times$ 6 mm is placed 41 cm downstream from the cell to collimate the molecular beam and to reduce the remaining $^4$He buffer-gas. The molecules interact with the lasers 50 cm from the cell where the in-vacuum RF coils are in place. After the interaction, the molecules travel 30 cm farther and are detected by fluorescence with an electron multiplying charge-coupled device (EMCCD) camera. All lasers include $X(v=0,1) - A(v'=0)$ transitions and their hyperfine splittings.}
\label{fig1} 
\end{center}
\end{figure}

A schematic of the apparatus is depicted in figure \ref{fig1}.
In our experiment, a cryogenic two-stage buffer-gas beam source is used to produce a cold CaF molecular beam \cite{Lu2011, Hutzler2012, Hemmerling2014, hemmerling2016}. 
CaF molecules are generated by ablating a metallic Ca target in an SF$_6$ ice environment within a copper cell, similar to the work in reference \cite{ImperialBeam}.
The molecules are cooled through collisions with He gas in a cell held at a temperature of around 2 K.
Cold molecules and atomic He exit the cell aperture, forming a beam with a mean forward velocity of 110 $\pm$ 10 m/s, and velocity spread of 40 m/s.
To reduce the amount of buffer gas reaching the interaction region and to collimate the molecular beam, a 6 mm $\times$ 6 mm aperture is placed 41 cm from the cell, 9 cm before the molecules interact with the RF-MOT magnetic fields and corresponding transverse laser beams. 
In these experiments only a single MOT beam -- transverse to the molecular beam -- is used to quantitatively study the RF magneto-optical force in 1D \cite{Hummon2013}.
After molecules pass through the interaction region, they travel about 30 cm farther where they are detected on an EMCCD using fluorescence. 

\begin{figure}
\begin{center}
\includegraphics[width=0.9\textwidth]{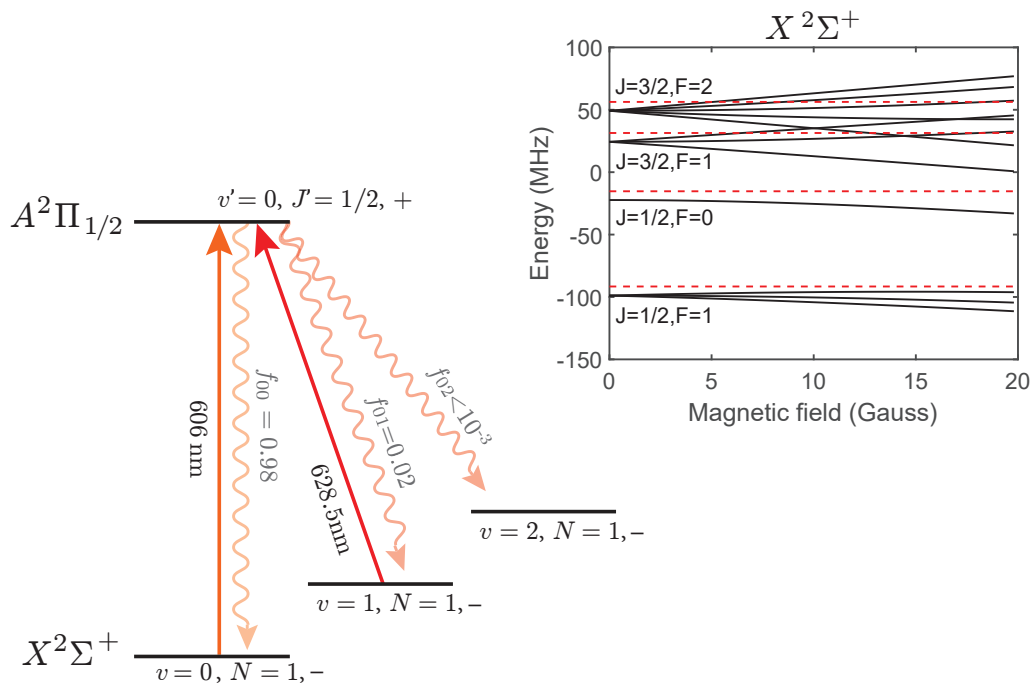}
\caption{Level diagram for CaF molecules. Only relevant levels are shown. Quantum numbers for vibration, rotation, electron's total angular momentum, and the atom's total angular momentum are $v$, $N$, $J$, and $F$ respectively. The parity of the state is denoted $\pm$. The straight lines indicate lasers used in the experiment and wavy lines show the decay from the excited states. FCFs ($f_{ij}$) are also shown. The inset shows the hyperfine states and their Zeeman shift in the ground state $X(v=0, N=1)$.  The red-detuned laser frequencies are shown in red broken lines in the inset. The hyperfine splitting and the Zeeman shift of the excited state $A(v'=0, J'=1/2)$ is negligible \cite{Devlin2015}.}
\label{fig2}
\end{center}
\end{figure}

The relevant energy levels of CaF are shown in figure \ref{fig2}. 
The optical transition in use is from the electronic ground state $X$ to the first electronic excited state $A$.
The measured FCF for $X(v=0) - A(v'=0)$ is 0.987 \cite{Wall2008} and one vibrational repump laser from the $v=1$ state is used, allowing up to 1000 photons to be scattered before pumping of the molecules to the $v=2$ state \cite{Pelegrini2005}.
Loss to other rotational states is prevented by addressing a $J$ to $J'=J-1$ transition \cite{Stuhl2008}. 

There are 4 hyperfine states that are spaced a few tens of MHz in the $X(N=1)$ state manifold due to coupling between the electron's spin $S=1/2$ and fluorine's nuclear spin $I = 1/2$.
This hyperfine splitting weakens the magneto-optical force in two ways: reduction of the scattering rate and level-crossing in weak magnetic fields at around 10 Gauss (figure \ref{fig2} inset).
All states are individually addressed through frequency sidebands placed on the lasers.
One of the hyperfine states ($J=1/2,F=1$) has an opposite sign of the $g$ factor and the polarization of the lasers addressing this state has opposite handedness relative to the other states.
The intensity of the $X(v=0) - A(v'=0)$ lasers for all hyperfine states is 150 mW/cm$^2$ in total, and the intensity of the repump lasers for the $X(v=1) - A(v'=0)$ transition is 100 mW/cm$^2$. 
The lasers are Gaussian beams with $1/e^2$ diameter of 7 mm. 
The powers among the hyperfine states are weighted by the degeneracy of the state. 
This is due to the fact that scattering rate for the MOT depends on the laser intensity for individual states as \cite{Norrgard2016a}
\begin{equation}
	R_{\textrm{sc}} = \Gamma \frac{n_{e}}{(n_{g}+n_{e})+2 \sum_{j=1}^{n_{g}}(1+4\Delta_{j}^{2} / \Gamma^{2}) I_{\textrm{sat,}j}/I_{j}},
	\label{eq_scattering}
\end{equation}
where $\Gamma$ is a linewidth of the excited states, $n_g$($n_e$) is the number of involved states in the ground(excited) state manifold, $\Delta_j$ is a laser detuning for the $j$ state, $I_{\textrm{sat},j}$ is a two-level saturation intensity for the $j$ state (4.87 mW/cm$^2$ and 4.37 mW/cm$^2$ for the $v=0$ and $v=1$ states respectively), and $I_j$ is laser intensity for the $j$ state.
The experimental parameters result in the scattering rate $R_{\textrm{sc}} \sim 2 \pi \times 0.43$ MHz from equation \ref{eq_scattering}.
The hyperfine splitting of the excited state $A$ is unresolved \cite{Wall2008}.

\begin{figure}
\begin{center}
\includegraphics[width=0.6\textwidth]{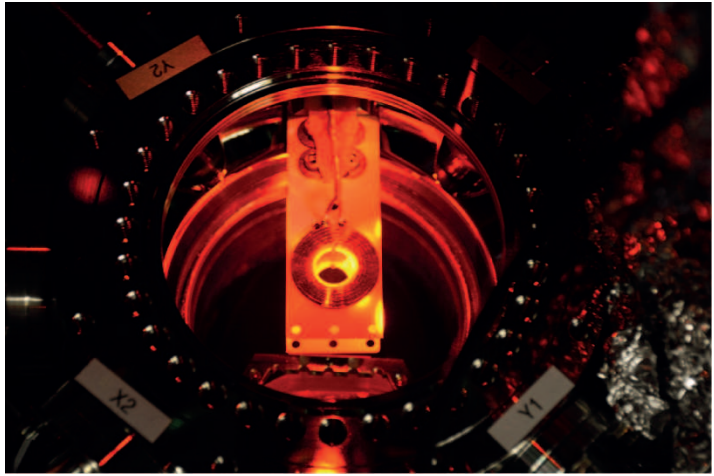}
\caption{RF MOT coils inside the vacuum chamber. The picture was taken before the inside was blackened with a non conductive, low outgassing paint (MH2200 from Alion) to reduce the background scattered light.}
\label{coilsphoto}
\end{center}
\end{figure}

One additional complication of a molecular MOT compared to an atomic MOT is that the number of involved ground magnetic substates is greater than that of the excited states. 
Because of this, some of the states become dark states of the confining lasers. 
The only way molecules could get out those states in a normal MOT would be scattering anti-confining photons, destroying the MOT effect.
RF MOTs \cite{Hummon2013, Norrgard2016a} solve this problem by actively switching the magnetic field gradient and the polarization of the light synchronously at a rate comparable to the excited state's lifetime.
With this scheme, molecules in all states predominantly scatter photons that lead to confinement and can be trapped.
To switch the needed magnetic fields at a rate on the order of MHz, the coils are made small and internal to the vacuum chamber (figure \ref{coilsphoto}).
The coils are made out of a 0.85 mm thick copper sheet cut to a spiral coil shape with wire width of 1 mm on a 1.5 mm thick alumina substrate .
The inner diameter of the coils is about 15 mm and there are 6 turns for each coil. 
There are 4 coils total, 2 on the upper board and 2 on the lower board -- with the boards spaced by 16 mm. 

Coils on the same board are wired in a Helmholtz configuration, while the two boards together provide an anti-Helmholtz field in the space between them. 
Alumina was chosen due to its low outgassing rate (necessary for the UHV environment of the MOT) and good thermal conductivity (necessary to carry out the substantial heating produced when switching the coils at high frequencies). 
The alumina boards are mounted on an aluminum block, which is itself mounted to a copper block feedthrough. 
On the air-side of the feedthrough, we put a passive heatsink to further increase the cooling power. 
With this setup, we observe only a few degrees of temperature increase of the block feedthrough when running the coils in RF mode at 830 kHz.
The coil assembly is connected to a resonant circuit outside of the chamber.
With 1 Amp of current run through the coils, a quadrupole magnetic field of about 7.3 (3.5) Gauss/cm is produced in the axial (radial) direction.

\begin{figure}
\begin{center}
\includegraphics[width=0.8\textwidth]{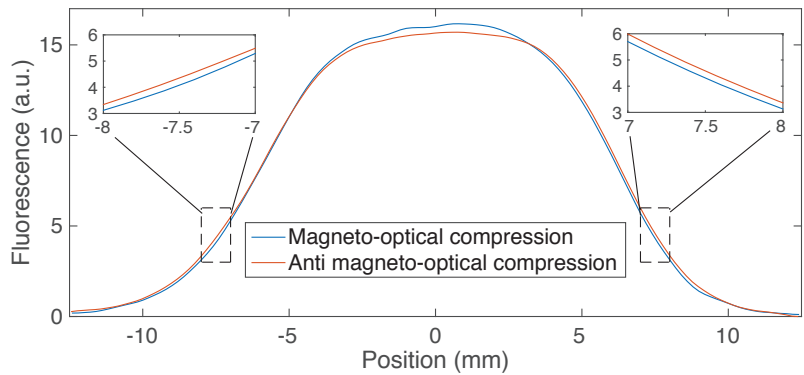}
\caption{Typical beam data with red-detuned lasers. The transverse beam shape after (anti) magneto-optical compression is depicted as a (red) blue line. The two configurations are achieved by changing the phase of the magnetic fields by 180 degrees. The integrated area of the beam is conserved for the both cases.}
\label{fig4}
\end{center}
\end{figure}

The molecules are detected by collecting fluorescent photons using an EMCCD camera. 
The excitation lasers ($X(v=0,1) - A(v'=0)$) cross the molecular beam at 90 degrees in the detection region, 80 cm downstream from the source.
The measured width of the molecular beam is a convolution of  the transverse velocity of the beam and the initial beam spread.
We analyze this molecular beam width to infer the compression and cooling effect from the magneto-optical force.

\section{Results and Discussion}

\begin{figure}
\begin{center}
\subfloat[With red-detuned lasers (-7 MHz).]{\includegraphics[width=0.49\textwidth]{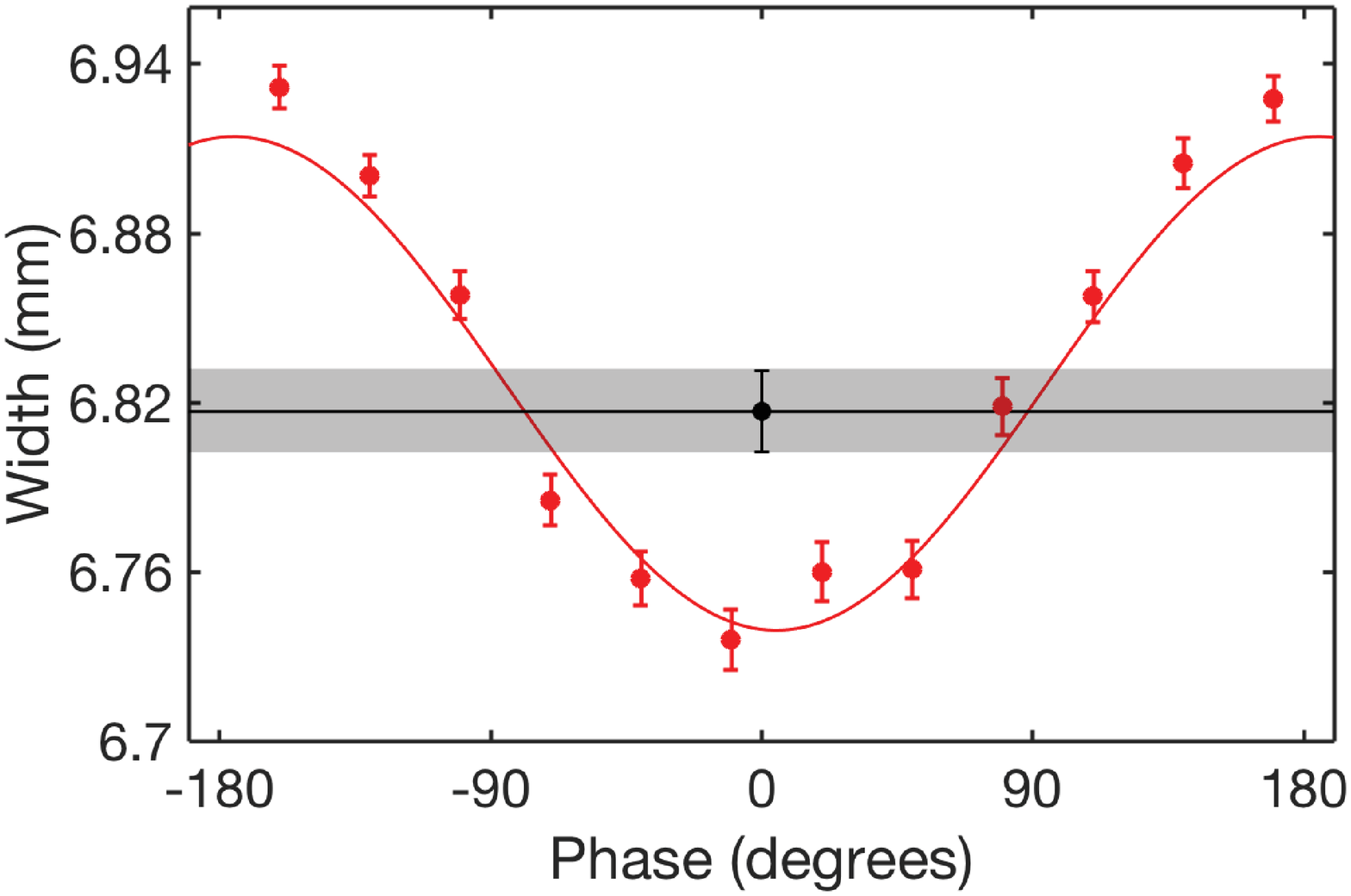}\label{fig5a}}
\subfloat[With blue-detuned lasers (+7 MHz).]{\includegraphics[width=0.49\textwidth]{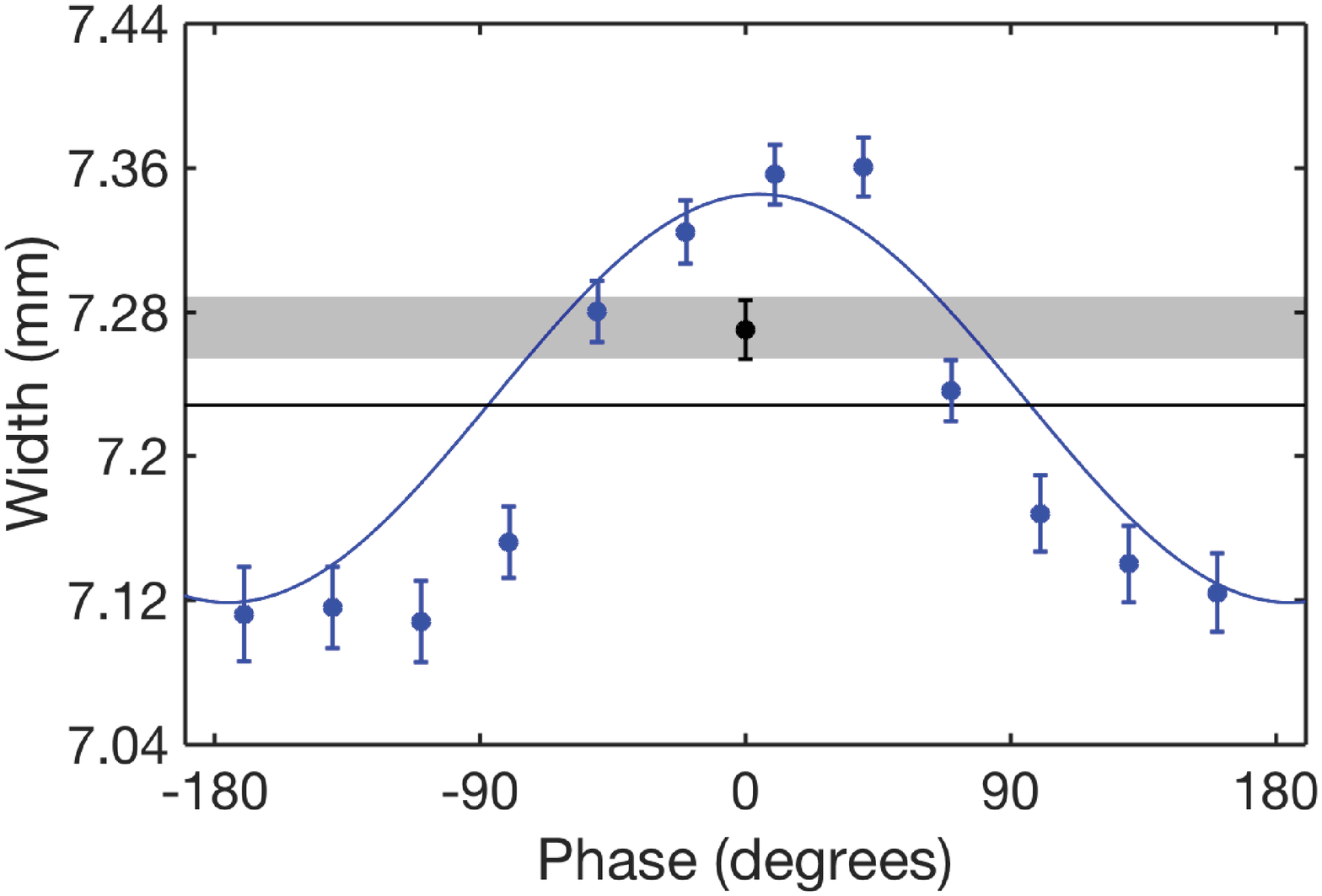}\label{fig5b}}
\end{center}
\caption{Molecular beamwidth are plotted as a function of the relative phase between the magnetic field and the polarization of the lasers: red dots in (a) for the $X(v=0) - A(v=0)$ laser detuning of -7 MHz, and blue dots in (b) for the $X(v=0) - A(v=0)$ laser detuning of +7 MHz. The black dot in each figure indicates the beamwidth only with Doppler effect taken without the magnetic field for each detuning. Gray area is a guide to eyes, indicating the $\pm 1\sigma$ confidence region of the Doppler width. Error bars are one-standard-deviation statistical uncertainties from fits to beam profiles. Lines show the Monte-Carlo simulation results for each case.}
\label{fig5}
\end{figure}

Figures \ref{fig4} and \ref{fig5} show the compression of the molecular beam using the magneto-optical force. 
Typical molecular beam signals with red-detuned $X(v=0) - A(v'=0)$ lasers when the magnetic field and the light polarization are in phase and out of phase are depicted in figure \ref{fig4}. 
The magneto-optical compression narrows the molecular beam when the phase of the magnetic field and the light polarization are matched.
In the opposite condition with the relative phase of 180 degrees, the molecular beam experiences an anti-confining force and its width is increased. 
Each data point in figure \ref{fig5} is the average of 350 measurements (molecular pulses) for the red-detuned magneto-optical compression and Doppler cooling (figure \ref{fig5a}), and 200(800) measurements for the blue-detuned magneto-optical compression (Doppler heating) (figure \ref{fig5b}). 
The relative phase between the polarization and the magnetic field is changed between each data point. 
The measured beam profiles are fit using super-Gaussian functions of order 4 with their amplitudes, beamwidths, and centers as free parameters.
Error bars indicate one-standard-deviation statistical uncertainties from fits to beam profiles.
In the absence of the magnetic field, Doppler cooling (heating) is observed with red (blue)-detuning of lasers. 
The direction of the magneto-optical force is seen to reverse with blue-detuning of the lasers compared to the case with red-detuning as expected.

\begin{figure}
\begin{center}
\includegraphics[width=0.65\textwidth]{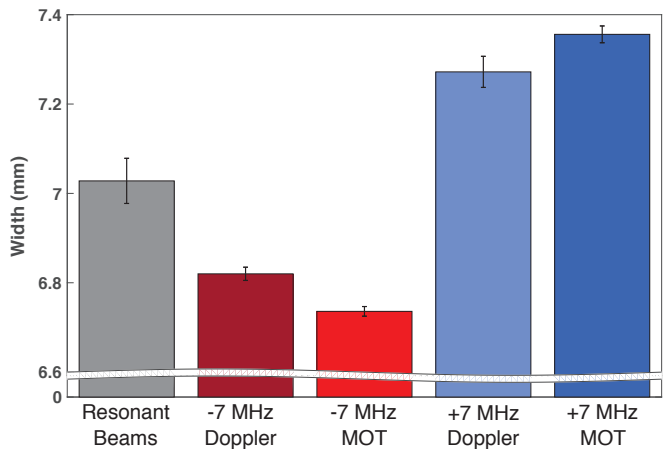}
\caption{Summary of the molecular beamwidth for each case. Both magnetic fields and detuned lasers are applied for the ``MOT'' cases, while no magnetic fields are applied for the ``Doppler'' cases. Molecular beamwidth after the molecules interacting with resonant lasers is shown for comparison. The error bars indicate one-standard-deviation statistical uncertainties from fits to beam profiles.}
\label{fig6}
\end{center}
\end{figure}

We have created Monte-Carlo simulations of the RF-MOT similar to that described in \cite{Hummon2013}. 
The results of these simulations are shown in figure \ref{fig5} as solid lines with scattering of 150 photons, corresponding to a photon scattering rate of $2\pi \times$ 0.4 MHz.
This agrees with the expected scattering rate based on the measured laser powers and transition strengths.
Due to the asymmetric hyperfine structure of CaF, a Monte-Carlo simulation with only one effective detuning gives good agreement for the red-detuning, while results are in worse agreement for the blue-detuned case. 
The widths of the molecular beam in the detection region for different conditions are summarized in figure \ref{fig6}.

At a detuning of 7 MHz, the optimal compression is achieved with a magnetic field gradient of 7.3 Gauss/cm.
Several other field gradients were tested but the compression effects are significantly reduced compared to the optimal gradient.
This is due to the small hyperfine splitting of CaF.
Specifically, in a magnetic field greater than 5 Gauss, the energy difference between magnetic sublevels of different hyperfine states approaches the natural linewidth. 
This increases the probability of scattering anti-confining photons, and thus weakens the compression effect.

The 1D magneto-optical compression demonstrated here provides us with an improved understanding of the 3D MOT for CaF. 
Using the number of scattered photons in 1D magneto-optical compression, simple calculations estimate a maximum (on-axis, perfect conditions) capture velocity of about 13 m/s for the 3D MOT in this experimental setup with our current geometry and a fairly generic laser configuration for the $X-A$ transition.
We have also performed a full 3D MOT Monte-Carlo simulation of trap loading for both on-axis and the more realistic set of molecular trajectories that include off-axis molecules.
Considering the limiting case of molecules entering the MOT region only directly on-axis with the zero magnetic field point of the MOT, the capture velocity is seen in the simulation to be 10 m/s, which agrees with the simple calculation. 
However, since the capture velocity decreases as the molecules travel off axis, we find that the overall capture velocity is about half of the ideal on-axis estimation, resulting in an effective capture velocity (simulating the molecular beam as a whole, including off-axis trajectories) of about 5 m/s for the molecular beam as a whole.
Based on the white-light slowing demonstrated previously in our group \cite{hemmerling2016}, significant numbers of molecules with velocities smaller than this capture velocity should be achievable with current techniques.

\section{Conclusion}
Magneto-optical compression of a buffer-gas cooled CaF beam has been achieved. 
By scattering 150 photons during the time molecules spend in the RF MOT region (about 60 $\mu$s), the beam is compressed, in good agreement with our first principles model.
From the demonstrated molecular beam compression, we estimate a RF MOT capture velocity of 5 m/s for CaF.

This work provides a deeper understanding of the magneto-optical forces on CaF molecules and guides the necessary experimental conditions for effective loading of molecular MOTs.
It further demonstrates the generality of the RF MOT scheme for molecules by adding an additional entry to the list of molecules on which magneto-optical forces have been demonstrated.

\ack
This work was supported by the ARO, the CUA, and the NSF. BLA is supported by the National Science Foundation Graduate Research Fellowship under NSF Grant No. DGE1144152.

\section*{References}
 
\bibliography{References}
\bibliographystyle{iopart-num}

\end{document}